\documentclass[psfig]{article}
\usepackage{amssymb}

\newcommand{\be}{\begin{equation}}
\newcommand{\ee}{\end{equation}}
\newcommand{\bea}{\begin{eqnarray}}
\newcommand{\eea}{\end{eqnarray}}

\title{Relativistic Burgers and Nonlinear Schr\"odinger Equations}

\author{Oktay K. PASHAEV
\\ \\
Department of Mathematics,
Izmir Institute of Technology \\
Urla-Izmir, 35430, Turkey}

\begin{document}
\maketitle

\begin{abstract}
Relativistic complex Burgers-Schr\"odinger and Nonliner Schr\"odinger equations are constructed.
In the non-relativistic limit they reduce to the standard Burgers and NLS equations respectively and
are integrable at any order of relativistic corrections.

\end{abstract}

\section{General Burgers-Schr\"odinger Hierarchy }
The relativistic linear Schr\"odinger equation has been discussed at the early years of quantum mechanics but
was dismissed promptly by the Klein-Gordon and the Dirac equations.
 Recently, relativistic versions of the Schr\"odinger
equation have been considered in
the study of relativistic quark-antiquark bound states \cite{BS}, and
gravitational collapse of a boson star \cite{FL}. A
nonlinear version of the model has appeared in the form of semi-relativistic
Hartree-Fock equation \cite{CO}. But none of those models is
known to be integrable. In the present paper we construct an
integrable relativistic nonlinear Schr\"odinger equation, preserving integrability at any
order of $1/c$ approximation.

We start from the Schr\"odinger equation in 1 + 1 dimensions

\begin{equation}
i\hbar \frac{\partial \Psi}{\partial t} = {\cal H}(P_1) \Psi \label{Schrodinger}
\end{equation}
for a free particle with classical dispersion of the general analytic form $E = E(p)$.  Here
$P_0 = i\hbar \frac{\partial}{\partial t}$ and $P_1 = -i\hbar \frac{\partial}{\partial x}$
are operators of the time and space translations respectively, commuting with
the Schr\"odinger operator $S = i\hbar \frac{\partial }{\partial t} - {\cal H}(P_1)$: $[P_\mu, S] = 0$, $\mu = 0,1$. The general boost operator,
defined as
$K = x - t {\cal H}'(P_1)$,
is also commuting with $S$, $[K,S] = 0$. Commuting it with space and time translations we have the algebra
of symmetry operators
\begin{equation}
[P_0, P_1] = 0,\,\,\,[P_0, K] = -\hbar {\cal H}'(P_1),\,\,\,[P_1,K] = - i\hbar \,.
\end{equation}
Then, if $\Psi$ is a solution of (\ref{Schrodinger}) and $W$ is an operator from this algebra, so that $[W, S] = 0$,  then $S \Psi$ is
also solution of  (\ref{Schrodinger}).

For a given classical dispersion $E = E(p)$, $E_0 \equiv E(0)$ we define $E$-polynomials $H^{(E)}_n (x,t)$ by the generating function
\begin{equation}
e^{\frac{i}{\hbar}( p x - (E(p)- E_0) t)} = \sum^\infty_{n=0} \left(\frac{i}{\hbar}\right)^n \frac{p^n}{n!} H^{(E)}_n(x,t)\,.
\end{equation}
It is equivalent to
\begin{equation}
H^{(E)}_n(x,t) = e^{-\frac{i}{\hbar}({\cal H}(-i\hbar \partial /\partial x)- E_0) t}\, x^n
\end{equation}
so that $H^{(E)}_n(x,t)$ is a solution of
\begin{equation}
i\hbar \frac{\partial }{\partial t}H^{(E)}_n(x,t) = ({\cal H}-E_0) H^{(E)}_n(x,t) \label{Schrodinger1}
\end{equation}
with the initial value $H_n^{(E)}(x,0) = x^n$.
From commutativity $[S,K] = 0$, time evolution of the operator $K$ satisfies
\begin{equation}
i\hbar \frac{\partial K}{\partial t} = [{\cal H}, K]
\end{equation}
and has the form
$K(t) = e^{-\frac{i}{\hbar}{\cal H} t}K(0)\, e^{\frac{i}{\hbar}{\cal H} t} = e^{-\frac{i}{\hbar}{\cal H} t} x \,e^{\frac{i}{\hbar}{\cal H} t}$.
Then as follows, operator $K$ generates the infinite hierarchy of polynomials according to
\begin{equation}
K H_n(x,t) = K e^{-\frac{i}{\hbar}({\cal H}-E_0) t}x^n =  e^{-\frac{i}{\hbar}({\cal H}-E_0) t}x^{n+1} = H_{n+1}(x,t)
\end{equation}

 \subsection{Non-relativistic Schr\"odinger equation}

 The non-relativistic dispersion $E(p) = p^2/2m$ implies the Hamiltonian operator
 \begin{equation}{\cal H} = -\frac{\hbar^2}{2m} \frac{\partial^2}{\partial x^2}\label{NH}\end{equation}
 and the Gallilean boost operator
 \begin{equation}
 K = x + i t \frac{\hbar}{m} \frac{\partial}{\partial x} \label{galileanboost}.
 \end{equation}
 From the generating function
\begin{equation}
e^{\frac{i}{\hbar}( p x - \frac{p^2}{2m} t)} = \sum^\infty_{n=0} \left(\frac{i}{\hbar}\right)^n \frac{p^n}{n!} H^{(S)}_n(x,t)
\end{equation}
we have the Schr\"odinger polynomials
\begin{equation}
H^{(S)}_n(x,t) = e^{\frac{i}{\hbar}t \frac{\hbar^2}{2m} \frac{\partial^2}{\partial x^2}} x^n \,.
\end{equation}
If $H^{(KF)}_n(x,t) = exp (t \frac{d^2}{dt^2}) x^n$ is the Kampe de Feriet polynomial, then
$H^{(S)}_n (x,t) = H^{KF} (x, \frac{i\hbar}{2m}t)$
or in terms of the Hermit polynomial
\begin{equation}
H^{(S)}(x,t) = \left( -\frac{i\hbar}{2m}t\right)^{n/2} H_n \left(\frac{x}{\sqrt{-2i\hbar t/m}}\right)\,.
\end{equation}

\subsection{Semi-relativistic Schr\"odinger equation}

The relativistic dispersion $E(p) = \sqrt{m^2 c^4 + c^2 p^2}$ implies the Hamiltonian
\begin{equation}
{\cal H} = mc^2 \sqrt{1 - \frac{\hbar^2}{m^2 c^2}\frac{\partial^2}{\partial x^2}}\label{SRH}
\end{equation}
and the semi-relativistic boost operator
\begin{equation}
K = x + \frac{i\hbar}{m} t \frac{\frac{\partial}{\partial x}}{\sqrt{1 - \frac{\hbar^2}{m^2 c^2}\frac{\partial^2}{\partial x^2}}}.
\end{equation}
In the non-relativistic limit, when $c \rightarrow \infty$, it reduces to the Galilean boost (\ref{galileanboost}).
The generating function in the form of relativistic plane wave
\begin{equation}
e^{\frac{i}{\hbar}( p x - (\sqrt{m^2 c^4 + c^2 p^2} - mc^2) t)} = \sum^\infty_{n=0} \left(\frac{i}{\hbar}\right)^n \frac{p^n}{n!} H^{(SRS)}_n(x,t)
\end{equation}
then gives  semi-relativistic polynomials
\begin{equation}
H^{SRS}_n(x,t) = e^{-\frac{i}{\hbar}mc^2 t (\sqrt{1 - \frac{\hbar^2}{m^2 c^2}\frac{\partial^2}{\partial x^2}} - 1)} x^n\,.
\end{equation}
In the non-relativistic limit $H^{SRS}_n \rightarrow H^S_{n} $. The first three polynomials coincide exactly with the Schr\"odinger polynomials
$H^{SRS}_1 (x,t) = x$, $H^{(SRS)}_2(x,t) = x^2 + i \frac{\hbar}{m} t$, $H^{(SRS)}_3 = x^3 + i\frac{\hbar}{m} 3 x t$,
while starting from the fourth one
\begin{equation}
H^{(SRS)}_4 (x,t)= x^4 + i\frac{\hbar}{m} 6 x^2 t - \frac{\hbar^2}{m^2} 3 t^2 + i\frac{\hbar^3}{m^3 c^2}3t
\end{equation}
we have relativistic corrections of order $1/c^2$. For complex valued space coordinate $x$, as it appears in 2+1 dimensional  Chern-Simons theory
\cite{PG}, zeros of these polynomials describe a motion
of point vortices in the plane. Equations of motion for $N$ vortices are (k = 1,...,N):
\begin{equation}
\dot x_k = \frac{i}{\hbar} Res_{|x=x_k} {\cal H}\left(\frac{\hbar}{i}\left(\frac{\partial}{\partial x} + \sum_{l=1}^N \frac{1}{x-x_l}\right)\right)\cdot 1
\end{equation}

\subsection{Relativistic Burgers-Schr\"odinger Equation}

 Using Schr\"odinger's log $\Psi$ transform \cite{S}, $\Psi = e^{\ln \Psi}$, and identity
 \begin{equation}
 e^{- \ln \Psi} \frac{\partial^n}{\partial x^n} e^{\ln \Psi} = \left(\frac{\partial}{\partial x} + \frac{\partial \ln \Psi}{\partial x} \right)^n \cdot 1
\end{equation}
the Schr\"odinger equation (\ref{Schrodinger}) can be rewritten in the form
\begin{equation}
i\hbar \frac{\partial }{\partial t} \ln \Psi = {\cal H} \left( - i\hbar \left( \frac{\partial}{\partial x} + \frac{\partial \ln \Psi}{\partial x}\right) \right)\cdot 1 \label{schrodinger3}
\end{equation}
For complex function $\Psi = e^{\frac{i}{\hbar}F} = e^{R + \frac{i}{\hbar} S}$ we
introduce a new complex function  $V = - i \frac{\hbar}{m} \frac{\partial}{\partial x} \ln \Psi = \frac{1}{m}\frac{\partial}{\partial x}F$,   with dimension of velocity.
 Then $F = S - i\hbar R$, is the complex potential, real and imaginary part of which are the  classical and quantum velocities
$V = V_c + i V_q = \frac{1}{m} S_x - i\frac{\hbar}{m}R_x$.
Hence (\ref{schrodinger3}) becomes of the complex Hamilton-Jacobi form (quantum Hamilton-Jacobi equation)
\begin{equation}
\frac{\partial F}{\partial t} + {\cal H}\left(-i\hbar \frac{\partial}{\partial x} + F_x\right)\cdot 1 = 0 \label{CHJ}
\end{equation}
In the classical (dispersionless) limit $\hbar \rightarrow 0$, the quantum velocity $V_q$ vanishes and complex potential $F$ reduces to the
real velocity potential $S$, playing the role of Hamilton's principal function. In this case (\ref{CHJ}) becomes the classical Hamilton-Jacobi equation
$\frac{\partial S}{\partial t} + {\cal H}\left(\frac{\partial S}{\partial x}\right) = 0$.
By differentiation of (\ref{schrodinger3}) we have equation for the complex velocity
\begin{equation}
i\hbar \frac{\partial V}{\partial t} = -i \frac{\hbar}{m} \frac{\partial}{\partial x}\left[{\cal H} \left( - i\hbar  \frac{\partial}{\partial x} + m V \right)\cdot 1\right] \label{schrodinger4}
\end{equation}
This equation is the Madelung fluid type representation of the Schr\"odinger equation (\ref{Schrodinger}).
In the classical limit it gives the Newton equation
$ m \frac{\partial V_c}{\partial t} = - \frac{\partial }{\partial x}\left[ {\cal H}(m V_c)\right]$
or in the hydrodynamic type form
\begin{equation}
\frac{\partial V_c}{\partial t} + {\cal H}'(m V_c)\frac{\partial V_c}{\partial x} =0 \label{HF}
\end{equation}
which is just differentiation of the classical Hamilton-Jacobi equation. Equation (\ref{HF})
has implicit general solution as
$V_c(x,t) = f(x - {\cal H}'(mV_c) t)$,
where $f$ is an arbitrary function, and it develops shock at a critical time when derivative
$(V_c)_x$ is blowing up.
\subsubsection{Non-relativistic Burgers-Schr\"odinger equation}

In this case the  Schr\"odinger equation (\ref{Schrodinger})
with non-relativistic Hamiltonian (\ref{NH}) is equivalent to the nonlinear equation for complex velocity V
\begin{equation}
i\hbar \frac{\partial V}{\partial t} = -\frac{\hbar^2}{2m}\frac{\partial^2 V}{\partial x^2} - i\hbar V \frac{\partial V}{\partial x} \label{NBS}
\end{equation}
which we call the Burgers-Schr\"odinger equation. In terms of the real and imaginary parts it gives the Madelung fluid
with density $\rho = e^R$ and velocity $V_c$.
In the classical limit it
reduces to the one real equation for classical velocity $V_c$, namely the ordinary dispersionless Burgers equation.

\subsubsection{Semi-relativistic Burgers-Schr\"odinger equation}

For  Hamiltonian (\ref{SRH}), the "Burgerization" procedure described above  gives
the
semi-relativistic Burgers-Schr\"odinger equation
\begin{equation}
\frac{1}{c}\frac{\partial V}{\partial  t} + c \frac{\partial}{\partial x}\left[\sqrt{1 +
\frac{1}{m^2 c^2}\left(-i\hbar \frac{\partial}{\partial x} + m V\right)^2} \cdot 1 \right] = 0\label{SRBS}
\end{equation}
In the non-relativistic limit it reduces to (\ref{NBS}) with relativistic corrections in the lowest order as
\begin{eqnarray}
i\hbar \frac{\partial V}{\partial t} = -\frac{\hbar^2}{2m}\frac{\partial^2 V}{\partial x^2} - i\hbar V \frac{\partial V}{\partial x} +
\frac{1}{8m^3 c^2}\left[-\hbar^4 V_{xxxx}  \right] \\
+ \frac{1}{8m^3 c^2}\left[-  i m \hbar^3 (10 V_x V_{xx} + 4 V V_{xxx} )+   m^2 \hbar^2 (12 V V_x^2 + 6 V^2 V_{xx}) + 4i m^3 \hbar V^3 V_x  \right]
\label{NCBS}
\end{eqnarray}
In the classical (dispersionless) limit (\ref{SRBS}) becomes equation of the hydrodynamic type
\begin{equation}
(V_c)_t + \frac{V_c}{\sqrt{1 + V_c^2/c^2}} (V_c)_x = 0 \label{DLSRBS}
\end{equation}
In the non-relativistic limit it reduces to the dispersionless Burgers equation with the lowest  relativistic correction
\begin{equation}
(V_c)_t + V_c (V_c)_x - \frac{1}{2c^2}V_c^3 (V_c)_x = 0
\end{equation}
The general implicit solution of (\ref{DLSRBS}) is
\begin{equation}
V_c(x,t) = f\left(x - \frac{V_c\, t}{\sqrt{1 + V_c^2/c^2}}\right)
\end{equation}
and it develops shock at a finite time.

\subsection{B\"acklund Transformation for Burgers-Schr\"odinger equation}

By the boost transformation of Section 1, from a given solution $\Psi_1$ of the Schr\"odinger equation
(\ref{Schrodinger}) we can generate another solution as
\begin{equation}
\Psi_2 = K \Psi_1 = \left[x - t {\cal H}'\left(-i\hbar \frac{\partial}{\partial x}\right)\right] \Psi_1\,.
\end{equation}
Using identity
\begin{equation}
\Psi^{-1} G\left(-i\hbar \frac{\partial}{\partial x}\right) \Psi = G\left(-i\hbar \frac{\partial}{\partial x} + m V\right)\cdot 1\end{equation}
for complex velocities $V_a = -i \frac{\hbar}{m} \ln \Psi_a$, $(a=1,2)$, we obtain the B\"acklund transformation
\begin{equation}
V_2 = V_1 - i \frac{\hbar}{m} \frac{\partial}{\partial x} \ln \left[x - t H'\left(-i\hbar \frac{\partial}{\partial x} + m V \right)\cdot 1\right]
\end{equation}

For the non-relativistic quantum mechanics (\ref{NH}) it gives complex B\"acklund transformation
\begin{equation}
V_2 = V_1 - i \frac{\hbar}{m} \frac{1 - (V_1)_x t}{x - V_1 t}
\end{equation}
for the Burgers-Schr\"odinger equation (\ref{NBS}).

For the semi-relativistic quantum mechanics (\ref{SRH}) we have the B\"acklund transformation of the form
\begin{equation}
V_2 = V_1 - i \frac{\hbar}{m} \frac{\partial}{\partial x} \ln \left( x - \frac{1}{\sqrt{1 + \frac{1}{m^2 c^2}
(-i\hbar \frac{\partial}{\partial x} + m V_1)^2}} V_1 t\right)
\end{equation}
It is worth to note that in the classical limit $\hbar \rightarrow 0$, $V \rightarrow V_c$ the above B\"acklund transformations  reduce to the trivial identity ${V_c}_1 = {V_c}_2$.

\section{Integrable General NLS Hierarchy}

In previous sections we studied the so called C-integrable relativistic Burgers-Schr\"odinger equation.
Now using the AKNS hierarchy for the NLS equation we are going to construct relativistic NLS.

\subsection{NLS hierarchy}

We consider the Zakharov-Shabat linear problem
\begin{equation}
\frac{\partial }{\partial x}\left(\begin{array}{c} v_1\\
v_2\end{array} \right) = \left( \begin{array}{cc} -
\frac{i}{2}p& -\kappa^2 \bar\psi \\ \psi &\frac{i}{2}p
\end{array}\right)\left(\begin{array}{c} v_1\\
v_2\end{array} \right) = J_1 \left(\begin{array}{c} v_1\\
v_2\end{array} \right) , \label{ZS11}\end{equation}
for the space
evolution, and the generalized AKNS problem \cite{AKNS}
\begin{equation}
\frac{\partial }{\partial t}\left(\begin{array}{c} v_1\\
v_2\end{array} \right) = \left( \begin{array}{cc} - i A&
-\kappa^2 \bar C \\ C & -i A
\end{array}\right)\left(\begin{array}{c} v_1\\
v_2\end{array} \right) = J_0 \left(\begin{array}{c} v_1\\
v_2\end{array} \right) , \label{ZS21}\end{equation} for the time
evolution, where for the real $A(x,t,p)$ and complex $C(x,t,p)$
functions, determined by the zero-curvature condition,
we substitute
$A_N = \sum_{n=0}^N A^{(n)} \left(-\frac{p}{2}\right)^n$, $C_N
= \sum_{n=0}^N C^{(n)} \left(-\frac{p}{2}\right)^n$.
It gives the evolution equation
$
\partial_{t_N} \psi = \partial_x C^{(0)} + 2i A^{(0)}\psi$
and $C^{(N)} = 0$, $A^{(N)} = a_N = const.$. We fix this constant so that $a_N = (-2)^{N-1}$.
Then we have the recurrence relations
$C^{(n)} = \frac{1}{2i} \partial_x C^{(n+1)} + A^{(n+1)}\psi$,
$\partial_x A^{(n)} = i \kappa^2 (\bar C^{(n)} \psi - C^{(n)} \bar\psi)$,
where $n = 0,1,2,...,N-1$. Integrating the last equation one has
\begin{equation}
A^{(n)} = -i \kappa^2 \int^x( \bar\psi C^{(n)}  - \psi\bar
C^{(n)} )\label{Aint}
\end{equation}
Substituting (\ref{Aint}) into recursion formula we find
\begin{equation}
\left( \begin{array}{c} C^{(n)}\\ \bar C^{(n)} \end{array}\right)
= -\frac{1}{2}\cal{R}\left( \begin{array}{c} C^{(n+1)}\\ \bar
C^{(n+1)} \end{array}\right)
\end{equation}
where ${\cal R}$ is the matrix integro-differential operator - the
recursion operator of the NLS hierarchy \cite{AKNS} -
\begin{equation} {\cal{R}} = i\sigma_3\left(\begin{array}{cccr}\partial_x+ 2\kappa^2 \psi
\int^x \bar\psi & -2\kappa^2 \psi \int^x \psi
\\ & \\-2\kappa^2 \bar\psi \int^x
\bar\psi&\partial_x+2\kappa^2\bar\psi \int^x \psi\end{array}
\right) \, \label{recursion}
\end{equation}
and $\sigma_3$ - the Pauli matrix.
Then we get
\begin{equation} i\sigma_3  \left (\begin{array}{clcr}\psi \\
\bar\psi \end{array} \right)_{t_{N}}= {\cal{R}}^{N} \left (\begin{array}{clcr} \psi\\
\bar\psi \end{array} \right) \label{NLShierarchy}\end{equation}
where $t_N$, $N = 1, 2, 3, ...$ is an infinite time hierarchy.
In the linear approximation, when $\kappa = 0$, the recursion
operator is just the momentum operator
${\cal{R}}_0 = i \sigma_3 \frac{\partial}{\partial x}$
and the NLS hierarchy (\ref{NLShierarchy}) becomes the linear
Schrodinger hierarchy
\begin{equation}
i\psi_{t_n} = i^n \partial^n_x \psi \,
\end{equation}
from Section 1.
The Madelung representation for this hierarchy, produced by the
complex Cole-Hopf transformation, is given by the complex
Burgers hierarchy \cite{PG}.

Every equation of hierarchy  (\ref{NLShierarchy}) is integrable. The linear problem for the
$N$-th equation is given by the Zakharov-Shabat problem (\ref{ZS11}) for the space part and
\begin{equation}
\frac{\partial }{\partial t_N}\left(\begin{array}{c} v_1\\
v_2\end{array} \right) = \left( \begin{array}{cc} - i A_N&
-\kappa^2 \bar C_N \\ C_N & -i A_N
\end{array}\right)\left(\begin{array}{c} v_1\\
v_2\end{array} \right) = J_{0_N} \left(\begin{array}{c} v_1\\
v_2\end{array} \right) , \label{ZSTN}\end{equation}
for the time part. Coefficient functions $C_N$ can be found conveniently 
as
\begin{equation}
\left( \begin{array}{c} C_N\\ \bar C_N
\end{array}\right) = \sum_{k=1}^{N} p^{N-k} {\cal{R}}^{k-1}\left( \begin{array}{c} \psi\\ \bar \psi
\end{array}\right) = (p^{N-1} + p^{N-2} {\cal{R}} + ... + {\cal{R}}^{N-1}) \left( \begin{array}{c} \psi\\ \bar \psi
\end{array}\right)
\end{equation}
To rewrite this expression in a compact form  we introduce notation of the  q-number
operator
\begin{equation}
1 + q + q^2 + ... + q^{N-1} \equiv [N]_q
\end{equation}
where $q$ is a linear operator. Hence, with operator $q \equiv
{\cal{R}}/p$ we have following finite Laurent form in the spectral
parameter $p$
\begin{equation}
1 + \frac{{\cal{R}}}{p} + \left(\frac{{\cal{R}}}{p}\right)^2 + ...
+ \left(\frac{{\cal{R}}}{p}\right)^{N-1} \equiv [N]_{{\cal{R}}/p}
\end{equation}
Then we have shortly
\begin{equation}
\left( \begin{array}{c} C_N\\ \bar C_N
\end{array}\right) = p^{N-1}[N]_{{\cal{R}}/p}\left( \begin{array}{c} \psi\\ \bar \psi
\end{array}\right)\label{CN}
\end{equation}
In a similar way
\begin{equation}
A_N = - \frac{p^N}{2} - i\kappa^2 p^{N-1}\left( \int^x \bar\psi, -
\int^x \psi \right) [N]_{{\cal{R}}/p}\left(
\begin{array}{c} \psi\\ \bar \psi
\end{array}\right)\label{AN}
\end{equation}
Equations (\ref{ZSTN}),(\ref{CN}) and (\ref{AN}) give the time part of the linear problem
(the Lax representation) for the N-th flow of NLS hierarchy (\ref{NLShierarchy}) in the q-calculus form.

\subsection{General NLS hierarchy equation}

For time $t$ determined by the formal series
$\partial_t = {\sum^\infty_{N=0}} E_N \partial_{t_{N}}$
where $E_N$ are arbitrary constants, the general NLS hierarchy
equation is
\begin{equation} i\sigma_3  \left (\begin{array}{clcr}\psi \\
\bar\psi \end{array} \right)_{t}= \left(E_0 + E_1 {\cal{R}} + ... + E_N {\cal{R}}^{N} +...\right)\left (\begin{array}{clcr} \psi\\
\bar\psi \end{array} \right) \label{GNLShierarchy}\end{equation}

Integrability of this equation is associated with
the Zakharov-Shabat problem (\ref{ZS11}) and the time evolution
\begin{equation}
J_0 = \sum_{N=0}^\infty E_N J_{0_N} = \left( \begin{array}{cc} -
i A& -\kappa^2 \bar C \\ C & -i A
\end{array}\right)
\end{equation}
where
\begin{equation}
\left( \begin{array}{c} C\\ \bar C
\end{array}\right)=
\sum_{N=0}^\infty E_N \left( \begin{array}{c} C_N\\ \bar C_N
\end{array}\right) = \sum_{N=1}^\infty E_N p^{N-1}[N]_{{\cal{R}}/p}\left( \begin{array}{c} \psi\\ \bar \psi
\end{array}\right)\label{CH}
\end{equation}
In the last equation we have used that for $N=0$, $C_0 = 0$. Then
we have
\begin{equation}
A = \sum_{N=0}^\infty E_N A_N = - \frac{1}{2}\sum_{N=0}^\infty E_N
p^N - i\kappa^2 \left( \int^x \bar\psi, - \int^x \psi \right)
\left(
\begin{array}{c} C\\ \bar C
\end{array}\right)\label{AH}
\end{equation}

The above equation (\ref{GNLShierarchy}) gives integrable
nonlinear extension of linear Schr\"dinger equation with general
analytic dispersion considered in Section 1. Let one considers the classical particle
system with the energy-momentum relation
$E(p) = E_0 + E_1 p + E_2 p^2 + ...$.
Then the corresponding time-dependent Schr\"odinger wave equation
is (\ref{Schrodinger})
where the Hamiltonian operator results from the standard
substitution for momentum $p \rightarrow -i \hbar
\frac{\partial}{\partial x}$ in the dispersion.
Equation (\ref{Schrodinger}) together with its complex conjugate can be
rewritten as
\begin{equation}
i \hbar \sigma_3 \frac{\partial}{\partial t}\left (\begin{array}{clcr}\psi \\
\bar\psi \end{array} \right) = H\left(-i \hbar \sigma_3
\frac{\partial}{\partial x}\right)\left (\begin{array}{clcr}\psi \\
\bar\psi \end{array} \right)\label{MSchr}
\end{equation}
The momentum operator here is just the recursion operator
in the linear approximation
${\cal{R}}_0 = i
\sigma_3 \frac{\partial}{\partial x}$. Hence, (\ref{MSchr}) is the linear Schr\"odinger equation with arbitrary
analytic dispersion.
The nonlinear integrable extension of this equation appears
as (\ref{GNLShierarchy}), which corresponds to the replacement
${\cal{R}}_0 \rightarrow {\cal{R}}$, ($\hbar = 1$), so that
\begin{equation} i\sigma_3  \left (\begin{array}{clcr}\psi \\
\bar\psi \end{array} \right)_{t}= H\left(\cal{R}\right)\left (\begin{array}{clcr} \psi\\
\bar\psi \end{array} \right) \label{GNLShierarchy1}\end{equation}
From this point of view, the standard substitution for classical
momentum $p \rightarrow -i \hbar \frac{\partial}{\partial x}$
or equivalently  $p \rightarrow -i \hbar
\sigma_3\frac{\partial}{\partial x} = {\cal{R}}_0$ for the
equation in spinor form, gives quantization in the form of the
linear Schr\"odinger equation. While substitution $p \rightarrow
{\cal{R}}$ gives "nonlinear quantization" and the nonlinear
Schr\"odinger hierarchy equation.

The related Lax representation for equation (\ref{GNLShierarchy1})
is given by (\ref{CH}), (\ref{AH}). By definition of
q-derivative
$D_q^{(\zeta)}f(\zeta) = \frac{f(q \zeta)-
f(\zeta)}{(q-1)\zeta}$
for operator $q = {\cal{R}}/p$, we have relation
$D_{{\cal{R}}/p}^{(p)} \zeta^N = [N]_{{\cal{R}}/p}\, p^{N-1}$.
Then equation (\ref{CH}) can be rewritten as
\begin{equation}
\left( \begin{array}{c} C\\ \bar C
\end{array}\right)=
 \sum_{N=1}^\infty E_N \, p^{N-1}[N]_{{\cal{R}}/p}\left( \begin{array}{c} \psi\\ \bar \psi
\end{array}\right)= \sum_{N=1}^\infty E_N D_{{\cal{R}}/p}^{(p)} \, p^N\left( \begin{array}{c} \psi\\ \bar \psi
\end{array}\right)
\label{CDH1}
\end{equation}
or using linearity of q-derivative  and analytic dispersion form
\begin{equation}
\left( \begin{array}{c} C\\ \bar C
\end{array}\right)=
 D_{{\cal{R}}/p}^{(p)}\sum_{N=0}^\infty E_N \, p^N\left(
\begin{array}{c} \psi\\ \bar \psi
\end{array}\right)= D_{{\cal{R}}/p}^{(p)}\, E(p)\left(
\begin{array}{c} \psi\\ \bar \psi
\end{array}\right)
\label{CDH2}
\end{equation}
Due to above definition  it gives simple formula
\begin{equation} \left( \begin{array}{c} C\\ \bar C
\end{array}\right)=
  \frac{E({\cal{R}})- E(p)}{{\cal{R}} - p}\left(
\begin{array}{c} \psi\\ \bar \psi
\end{array}\right) \label{CDH3}
\end{equation}
 Then for $A$ we obtain
\begin{equation}
A =  - \frac{1}{2}E(p) - i\kappa^2 \left( \int^x \bar\psi, -
\int^x \psi \right) \frac{E({\cal{R}})- E(p)}{{\cal{R}} - p}\left(
\begin{array}{c} \psi\\ \bar \psi
\end{array}\right)\label{AH3}
\end{equation}
Equations (\ref{CDH3}),(\ref{AH3}) give the Lax representation of
the general integrable NLS hierarchy model (\ref{GNLShierarchy1}) in a simple and compact form.
It is worth to note that special form of the dispersion $E=
E(p)$ is fixed by physical problem. In the next Section  we will discuss
the relativistic form of this dispersion and corresponding
semi-relativistic NLS equation.

\section{Semi-relativistic NLS} In Section 1.2 we have considered the relativistic dispersion relation 
$E(p) =
\sqrt{m^2c^4 + p^2c^2}$.  This may be used to construct a ``semi-relativistic" Schr\"odinger
equation with Hamiltonian (\ref{SRH}). Then,
combining two complex conjugate equations together, we have
\begin{equation} i\sigma_3  \left (\begin{array}{clcr}\psi \\
\bar\psi \end{array} \right)_{t}= mc^2 \sqrt{ 1 + \frac{1}{m^2 c^2}{\left(i\sigma_3 \frac{\partial}{\partial x}\right)}^{2}} \left (\begin{array}{clcr} \psi\\
\bar\psi \end{array} \right) \label{RelLS}\end{equation}
We like to emphasize that if $\psi$ describes the relativistic particle forward in time and with positive energy,
$\bar \psi$ corresponds to the backward time or to the negative energy. From this point of view equation (\ref{RelLS}) is complete since includes both states.

Following the general procedure described in previous Section
one may proceed further: by replacing the derivative
operator ${\cal R}_0 = i\sigma_3\frac{\partial}{\partial x}$ corresponding to linear momenta $p$
with the full recursion operator $\cal{R}$
(\ref{recursion}), one obtains an {\it integrable} relativistic
nonlinear Schr\"odinger equation

\begin{equation} i\sigma_3  \left (\begin{array}{clcr}\psi \\
\bar\psi \end{array} \right)_{t}= mc^2 \sqrt{ 1 + \frac{1}{m^2 c^2}{\cal{R}}^{2}} \left (\begin{array}{clcr} \psi\\
\bar\psi \end{array} \right) \label{RelNLS}\end{equation} where
the square root operator has meaning of the formal power series so
that
\begin{equation} i\sigma_3  \left (\begin{array}{clcr}\psi \\
\bar\psi \end{array} \right)_{t}= mc^2 \left(1 + \frac{1}{2 m^2 c^2}{\cal{R}}^{2} - \frac{1}{8 m^4 c^4}{\cal{R}}^{4} +
\frac{1}{16 m^6 c^6}{\cal{R}}^{6}\pm ...\right) \left (\begin{array}{clcr} \psi\\
\bar\psi \end{array} \right) \label{Rel1NLS}\end{equation}
For the above relativistic
dispersion and equation
(\ref{RelNLS}), we have the next linear problem
\begin{equation}
\frac{\partial }{\partial x}\left(\begin{array}{c} v_1\\
v_2\end{array} \right) = \left( \begin{array}{cc} -
\frac{i}{2}p& -\kappa^2 \bar\psi \\ \psi &\frac{i}{2}p
\end{array}\right)\left(\begin{array}{c} v_1\\
v_2\end{array} \right) , \label{ZS1RelNLS}\end{equation}

\begin{equation}
\frac{\partial }{\partial t}\left(\begin{array}{c} v_1\\
v_2\end{array} \right) =  \left(
\begin{array}{cc} - i A& -\kappa^2 \bar C \\ C & -i A
\end{array}\right)\left(\begin{array}{c} v_1\\
v_2\end{array} \right),\label{ZS2RelNLS}
\end{equation}
where
\begin{equation} \left( \begin{array}{c} C\\ \bar C
\end{array}\right)=
  \frac{\sqrt{m^2c^4 + {{\cal R}}^2c^2}- \sqrt{m^2c^4 + p^2c^2}}{{\cal{R}} - p}\left(
\begin{array}{c} \psi\\ \bar \psi
\end{array}\right) \label{CRel}
\end{equation}
\begin{equation}
A =  - \frac{1}{2}\sqrt{m^2c^4 + p^2c^2} - i\kappa^2 \left(
\int^x \bar\psi, - \int^x \psi \right) \frac{\sqrt{m^2c^4 + {{\cal
R}}^2c^2}- \sqrt{m^2c^4 + p^2c^2}}{{\cal{R}} - p}\left(
\begin{array}{c} \psi\\ \bar \psi
\end{array}\right)\label{ARel}
\end{equation}
and the spectral parameter $p$ has meaning of the classical
momentum.
 The model (\ref{RelNLS}),
 is an integrable nonlinear
Schrodinger equation with relativistic dispersion:
\begin{equation}
i \psi_t = m c^2 \sqrt{1 - \frac{1}{m^2
c^2}\frac{\partial^2}{\partial x^2}}\, \psi + F(\psi)
\end{equation}
where the nonlinearity expanded in $1/c^2$ is the infinite sum
\begin{eqnarray}
\lefteqn{F(\psi) = \frac{1}{2m}[-2\kappa^2 |\psi|^2 \psi]}
\nonumber
\\
& & - \frac{1}{8m^3 c^2} [2\kappa^2(2|\psi_x|^2\psi +
4|\psi|^2\psi_{xx} + \bar\psi_{xx}\psi^2 + 3\bar\psi \psi^2_x) +
6\kappa^4 |\psi|^4
\psi] + O(\frac{1}{c^4}) \nonumber
\end{eqnarray}

It is interesting to note that if we expand also the dispersion part in
$1/c^2$, then at every order of $1/c^2$  we get an integrable
system. It means that  we obtain integrable relativistic corrections
to the NLS equation at any order. From the known relativistic integrable models
like the Sine-Gordon or the Liouville equations, neither one
has this property. Finally we note that nonlinear relativistic equations considered in this paper are
distinct from those obtained in \cite{BS}-\cite{CO} and references therein. They 
might be useful in analyzing relativistic
corrections to solitons, Bose-Einstein condensates or other condensed matter systems with an effective equation of relativistic form.

\section{Acknowledgements}
This work partially supported by Izmir Institute of Technology, BAP Grant 2008IYTE25.

\end{document}